\documentclass[11pt]{article}
\usepackage[textwidth=480pt,textheight=680pt]{geometry}
\usepackage{amsmath}
\usepackage{amssymb}
\usepackage{bbold}
\usepackage{xspace}
\usepackage[bottom]{footmisc}
\usepackage{graphicx}
\usepackage{psfrag}
\usepackage{hyperref}
\usepackage[bf,small]{caption2}
\setcaptionwidth{.9\textwidth}
\makeatletter
\@addtoreset{equation}{section}

\makeatother
\usepackage[numbers]{natbib}

\newdimen\tableauside\tableauside=1.5ex   
\newdimen\tableaurule\tableaurule=.32pt   
\newdimen\tableaustep
\def\phantomhrule#1{\hbox{\vbox to0pt{\hrule height\tableaurule width#1\vss}}}
\def\phantomvrule#1{\vbox{\hbox to0pt{\vrule width\tableaurule height#1\hss}}}
\def\sqr{\vbox{%
  \phantomhrule\tableaustep
  \hbox{\phantomvrule\tableaustep\kern\tableaustep\phantomvrule\tableaustep}%
  \hbox{\vbox{\phantomhrule\tableauside}\kern-\tableaurule}}}
\def\squares#1{\hbox{\count0=#1\noindent\loop\sqr
  \advance\count0 by-1 \ifnum\count0>0\repeat}}
\def\tableau#1{\vcenter{\offinterlineskip
  \tableaustep=\tableauside\advance\tableaustep by-\tableaurule
  \kern\normallineskip\hbox
    {\kern\normallineskip\vbox
      {\gettableau#1 0 }%
     \kern\normallineskip\kern\tableaurule}%
  \kern\normallineskip\kern\tableaurule}}
\def\gettableau#1 {\ifnum#1=0\let\next=\null\else
  \squares{#1}\let\next=\gettableau\fi\next}

\DeclareMathOperator{\Tr}{Tr}
\newcommand{\bA}{B}
\newcommand{\bB}{C}

\begin{document}
\pagestyle{empty}
\begin{flushright}
AEI-2005-126\\
hep-th/0507178\\
\today
\end{flushright}
\vskip 5ex

\begin{center}
\begin{minipage}{.9\textwidth}
\lineskip 2ex
{\huge\bf Higher-derivative gauge field terms\\ in the M-theory action}\\[6ex]
{\large\bfseries Kasper Peeters, Jan Plefka and Steffen Stern}\\[5ex]
Max-Planck-Institut f\"ur Gravitationsphysik\\
Albert-Einstein-Institut\\
Am M\"uhlenberg 1\\
14476 Golm, Germany\\[3ex]
{\tt kasper.peeters},
{\tt jan.plefka},
{\tt steffen.stern@aei.mpg.de}
\vskip 7ex

\noindent {\bf Abstract:} We use superparticle vertex operator
correlators in the light-cone gauge to determine the~$(DF)^2\,R^2$
and~$(DF)^4$ terms in the M-theory effective action. Our results, when
compactified on a circle, reproduce terms in the type-IIA string
effective action obtained through string amplitude calculations.
\end{minipage}
\end{center}
\vfill
\newpage
\pagestyle{plain}
\hrule
\tableofcontents
\bigskip
\hrule
\medskip

\section{Introduction}

While a substantial effort has been spent on the computation of
effective actions for string theory, the situation is not as well
developed for M-theory due to our incomplete understanding of its
quantum structure. Just as in string theory, the effective action for
the massless modes (given by the graviton, the three-form and the
gravitino) consists of the lowest-order supergravity
action~\cite{crem1} plus an infinite tower of higher-derivative
terms. In string theory there exists a variety of methods which can be
used to determine this tower of terms. In the background field method
one couples the string to a background of supergravity
fields. Conformal invariance then demands the vanishing of a set of
\mbox{$\beta$--functionals} for these fields, which are interpreted as
equations of motion of the effective action. This method is clearly
not available in M-theory, as the membrane world-volume action does
not satisfy a similar constraint. An alternative method is to employ
supersymmetry constraints in order to determine the higher-derivative
action. However, determining them in practice along these lines has
proved to be hard, both in string theory and in M-theory~(see
e.g.~\cite{Peeters:2000qj,kas_towards,Cederwall:2004cg} for a
discussion of the present status of this programme).

A third method builds directly upon the S-matrix and extracts the
effective action from string scattering amplitudes. This method does
admit a certain generalisation to M-theory. From the field theory
point of view the higher order corrections are counterterms for the
non-renormalisable supergravity theory. Hence one way of determining
them is to perform loop computations in 11d supergravity, resulting in
explicit tensorial expressions of the supergravity fields with cutoff
dependent coefficients. These undetermined coefficients may often be
fixed by considering a compactification to ten or lower dimensions and
comparison to string theory~\cite{Green:1997as,Green:1999pu}.  In
order to actually perform these loop calculations it is advisable to
make use of a supersymmetric formalism where cancellations are
manifest. Although computations based on the covariant on-shell
superspace formalism~\cite{crem3,deWit:1998tk} have not yet appeared,
an efficient supersymmetric light-cone gauge has been developed by
Green, Gutperle and Kwon~\cite{Green:1999by}.  In this formalism,
one-loop supergravity amplitudes are described by a closed worldline
integral of the~11d superparticle with vertex operator insertions. The
zero-mode structure of the superparticle vertex operators then
dictates the vanishing of all one, two and three-point functions at
one-loop. The four-point amplitudes are of a very special form, as
they factorise into a scalar box diagram times the tensorial structure
which is completely determined by the fermionic zero mode integral over the
vertex operators. For four-gravitons this gives rise to the famous
$R^4$-term of quartic order in the Riemann tensor.
\medskip

In the present paper we determine all other superparticle amplitudes for
which the tensorial structure is again completely determined by a
fermionic zero mode integral. All four-point amplitudes are of this
protected nature. The bosonic ones which have not been computed so far
are those with four three-form fields or two three-form fields and two
gravitons. These lead to terms of the form $(DF)^4$ and $(DF)^2\, R^2$ in
the effective action, where $F={\rm d}C$ with $C$~being the M-theory
three-form potential.  For the determination of these amplitudes one
needs to compute a variety of Levi-Civita traces of~SO(9) Dirac
matrices coming from the fermionic zero-mode integral. The resulting
tensors generalise the $t_8$ tensor for the superstring defined
via~SO(8) Dirac matrices. In the present paper we evaluate all these
amplitudes and hence determine the structure of the M-theory effective
action in this sector. As a test of our results we also check that
they reduce to the known string theoretic quartic effective action
terms $(DH)^4$ and $(DH)^2\, R^2$ (with $H={\rm d}B$) in 10d computed by
Gross and Sloan~\cite{Gross:1987mw}.  Our final expressions may be
found in eqs.~\eqref{e:AR2DF2} and~\eqref{e:ADF4}.

As a matter of fact the superparticle vertex operator formalism may be
lifted to the~11d light-cone supermembrane where corresponding vertex
operators for the graviton, three-form and gravitino may be
defined~\cite{Dasgupta:2000df}. These vertices not only reduce to the
11d superparticle vertices once one shrinks the membrane space-sheet
to a point, but they also reduce to the type-IIA superstring vertices
under double-dimensional reduction. It turns out that scattering
amplitudes for the supermembrane theory may be defined in this
supersymmetric light-cone gauge in analogy with the string and
particle descriptions~\cite{Dasgupta:2000df,Plefka:2000gv}. The
tensorial structures of the four-point amplitudes then reduce to
precisely the same zero-mode integrals that one encounters in the
superparticle computation discussed above. Hence depending on one's
personal taste, one may consider our computation of the $(DF)^4$ and
$(DF)^2\,R^2$ terms to arise either from a superparticle or from a
supermembrane light-cone formalism.
\medskip

Let us end this introduction by comparing our method to alternative
ways in which higher-derivative terms in the M-theory effective action
can be computed.  Certainly the light-cone gauge has its drawbacks as
it is not able to compute all possible amplitudes in a generic
background. For example a term of the form~$(DF)^3\,R$ will always
require the contraction with an~11d epsilon-tensor due to the odd
total number of indices. Such terms are hence invisible in light-cone
gauge where~$R_{+mnp}=0$ and~$F_{+mnp}=0$.  It would therefore be
worth analysing our computations using the covariant formalism for~11d
loop-amplitudes as presented recently in~\cite{Anguelova:2004pg},
which makes use of previous work of Berkovits~\cite{Berkovits:2002uc}
for the covariant~11d supermembrane.

A completely different method for the computation of quartic terms in
the effective action was given by Deser and Seminara
in~\cite{Deser:1998jz,Deser:2000xz,Deser:2005kb}. They argued that one
may extract the form of local supergravity counterterms from the
nonlocal parts of the \emph{tree} level four-point amplitudes. Based
on this argument they have constructed $(DF)^4$, $(DF)^2\, R^2$ and
$(DF)^3\, R$ counterterms. While our findings agree with theirs for
the $(DF)^2\, R^2$ terms in the action and we also agree with their
nonlocal amplitudes for~$(DF)^4$, our local~$(DF)^4$ action differs
from the local form obtained in~\cite{Deser:2000xz}.

Finally one might wonder whether Matrix Theory~\cite{bank3} as a
contender for the microscopic definition of M-theory can also compute
these terms (potentially regularising the divergent prefactors). That
is, unfortunately, not the case, as determining these quantum
corrections directly from Matrix-theory presumably requires analytical
control on the large-$N$ form of the ground state wave
function. Investigations of supergraviton scattering amplitudes in
Matrix theory have proved to lead to inconsistencies at finite~$N$
once one probes beyond the leading supergravity approximation
\cite{Helling:1999js}. The leading one-loop Matrix Theory S-Matrix for
supergraviton scattering agrees with~11d supergravity, but is in fact
completely determined by supersymmetry~\cite{Nicolai:2000ht}
irrespective of the value of~$N$.  At two-loops in the~SU(3) model the
expected quantum correction of~$R^4$ is 
\emph{not} reproduced in the matrix model as was demonstrated
in~\cite{Helling:1999js}.

\section{Method}
\subsection{Membrane vertex operator correlators}

As explained in the introduction, we intend to determine
the~$(DF)^2\,R^2$ and $(DF)^4$ terms in the effective action through the
computation of superparticle or supermembrane scattering
amplitudes. Let us first recall the form of the vertex operators for
the physical states which feature in these calculations. Explicit
expressions for the membrane vertex operators in the light-cone gauge
have been constructed in~\cite{Dasgupta:2000df}.  They are given by
\begin{multline}
\label{e:Vh}
V_{h} = h_{ab} \Big[ DX^a DX^b - \{X^a, X^c\}\{X^b X^c\} - i \theta
\gamma^a \{X^b, \theta\} \\[1ex] 
- 2D X^a R^{bc} k_c - 6\{X^a, X^c\}
R^{bcd}k_{d} + 2 R^{ac} R^{bd} k_{c} k_{d}\Big] e^{-ik \cdot X}
\end{multline}
for the graviton and
\begin{multline} 
\label{e:VC}
V_{C} = -C_{abc} D X^{a} \{X^b, X^c \} e^{-ik \cdot X} \\[1ex] 
+ F_{abcd}
\left[ \left( DX^a - \tfrac{2}{3} R^{ae} k_e \right) R^{bcd} -
  \tfrac{1}{2} \{X^a, X^b\} R^{cd} -  \frac{1}{96} \{X^e, X^f\} \theta
  \gamma^{abcdef} \theta \right] e^{-ik \cdot X} 
\end{multline}
for the three-form.  The $X^a(\tau,\sigma_1,\sigma_2)$ and
$\theta_\alpha(\tau,\sigma_1, \sigma_2)$ denote the transverse
membrane embedding coordinates ($a=1,\ldots, 9$ and $\alpha=1,\ldots
,16$). Moreover $\{A,B\}:=\epsilon^{rs} \partial_rA\,\partial_s B$
where $\partial_r:=\partial/\partial \sigma_r$ is the Lie bracket of
area preserving diffeomorphisms under which the light-cone gauge fixed
supermembrane maintains a residual invariance~\cite{wit11}.  We also
have defined $DX:=\partial_t X+\{\omega,X\}$ with $\omega$ denoting
the gauge field of the area preserving diffeomorphisms.  In the above
the symbols $R^{abc}$ and $R^{ab}$ stand for the fermi bilinears
\begin{equation}
R^{abc} = \frac{1}{12} \theta \gamma^{abc} \theta\,,\qquad
R^{ab} = \frac{1}{4} \theta \gamma^{ab} \theta\,.
\end{equation}
Out of these, one can construct $n$-point 1-loop amplitudes
via~\cite{Plefka:2000gv,Dasgupta:2002iy}
\begin{equation}
{\cal A}_{\text{1-loop, $n$-point}} = 
  \int\!{\rm d}p^+{\rm d}^9 p_{\perp}\, \Tr(\Delta V_{1} \Delta V_{2} ... \Delta
  V_{n})\,.
\end{equation}
where the $V_{n}$ are vertex operators, $\Delta$ a propagator and the
trace goes over the Hilbert space.  For four-point amplitudes, almost
all terms in the expressions~\eqref{e:Vh} and~\eqref{e:VC} are,
however, irrelevant. This is because a non-zero amplitude requires the
saturation of a fermionic SO(9) integral, according to the identity
\begin{equation}
\Tr\big(\theta_{\alpha_1} \cdots \theta_{\alpha_N}\big)_{\theta} 
  = \delta_{N,16}\, \epsilon^{\alpha_1 \ldots \alpha_{16}}\,,
\end{equation}
Therefore, in computing four-point correlators, the only relevant
terms in the supermembrane vertex operators are those which are
already present in the superparticle vertex
operators~\cite{Green:1999by}. Those read
\begin{equation}
\label{e:Vsuperparticle}
\begin{aligned}
V_h &= 2\,h_{ab}\, R^{ac} R^{bd} k_c k_d \, e^{-ik\cdot X}\,,\\[1ex]
V_C &= -\tfrac{2}{3}\, F_{abcd} R^{ae} k_e R^{bcd}\,e^{-ik\cdot X}\,.
\end{aligned}
\end{equation}

As in string theory, the amplitudes will contain an overall
momentum-dependent factor arising from the correlator of plane-wave
exponentials. This factor is expected to exhibit poles corresponding
to massless field exchange graphs as well as regular terms
corresponding to contact terms. Lacking a firm understanding of these
integrals for the supermembrane we will not comment on the momentum
dependence any further. Progress on this correlator has been made by
passing back to a covariant description and working under the
assumption that the path integral reduces to a membrane zero-mode
winding sum while all quantum fluctuations cancel due to
supersymmetry~\cite{Pioline:2001jn,Kazhdan:2001nx,Pioline:2004xq}. For
the case of a three-torus compactification of M-theory this correlator
was studied in a three dimensional matrix theory description
\cite{Sugino:2001iq}, see also \cite{Bengtsson:2004nj} for a related
discussion.  We will here only assume that the light-cone correlator
yields a regular term, so that the amplitudes can be used to
determine~$(DF)^2\, R^2$ and $(DF)^4$ terms in the effective
action. It should be stressed however, that for a true supermembrane
theory reading of our results it remains to be shown that this scalar
correlator exists. Alternatively, one could take the point of view
that we are computing a superparticle correlator, in which case one
has to deal with the loop divergence like in~\cite{Green:1997as}.

Explicitly, one now finds that the four-point amplitudes with either
gravitons or three-form gauge fields consists of the above-mentioned
momentum dependent prefactor times the following tensor structures,
\begin{align} 
{\cal A}_{4h} &= 
   t_{16}^{a_1 a_2 \ldots a_{16}}\,
   R_{a_1 a_2 a_3 a_4} \cdots
   R_{ a_{13} a_{14} a_{15} a_{16}} \,,\\[1ex]
{\cal A}_{2h\,2C} &= 
   t_{18}^{a_1 a_2 \ldots a_{18}}\,
   D_{a_7}F_{a_{8} a_1 a_2 a_3}\,
   D_{a_{9}}F_{a_{10} a_4 a_5 a_6}\, 
   R_{a_{11} a_{12} a_{13} a_{14}}\,
   R_{a_{15} a_{16} a_{17} a_{18}} \,,\\[1ex]
{\cal A}_{4C} &= 
   t_{20}^{a_1 a_2 \ldots a_{20}}\,
   D_{a_{13}}F_{a_{14} a_1 a_2 a_3}\, 
   D_{a_{15}}F_{a_{16} a_4 a_5 a_6}\, 
   D_{a_{17}}F_{a_{18} a_7 a_8 a_9}\, 
   D_{a_{19}}F_{a_{20} a_{10} a_{11} a_{12}} \,.
\end{align}
We have here written Riemann tensors and derivatives of field
strengths instead of the linearised expressions in terms of
polarisation tensors and momenta, anticipating that these amplitudes
are directly responsible for the appearance of contact terms in the
effective action. The tensors~$t_{16}$, $t_{18}$ and $t_{20}$ are
generalisations of the well-known~$t_8$ tensor which appears in string
theory. Explicitly, these tensors are defined by
\begin{align}
t_{16}^{a_1 a_2 \ldots a_{16}} &:= 
  \epsilon^{\alpha_1 \ldots \alpha_{16}}\, 
  \gamma^{a_1 a_2}_{\alpha_1 \alpha_2}\, 
  \gamma^{a_3 a_4}_{\alpha_3 \alpha_4} \cdots
  \gamma^{a_{15} a_{16}}_{\alpha_{15} \alpha_{16}}\,,\\[1ex]
\label{e:t18def}
t_{18}^{a_1 a_2 \ldots a_{18}} &:= 
   \epsilon^{\alpha_1 \ldots \alpha_{16}}\,
   \gamma_{\alpha_1 \alpha_2}^{a_1 a_2 a_3}\, 
   \gamma_{\alpha_3 \alpha_4}^{a_4 a_5 a_6} \,
   \gamma_{\alpha_5 \alpha_6}^{ a_7 a_8 } \ldots 
   \gamma_{\alpha_{15} \alpha_{16}}^{ a_{17} a_{18}}\,,\\[1ex]
\label{e:t20def}
t_{20}^{a_1 a_2 \ldots a_{20}} &:= 
  \epsilon^{\alpha_1 \ldots \alpha_{16}} \,
  \gamma_{\alpha_1 \alpha_2}^{a_1 a_2 a_3} \cdots
  \gamma_{\alpha_7 \alpha_8}^{a_{10} a_{11} a_{12}} 
  \gamma_{\alpha_9 \alpha_{10}}^{a_{13} a_{14}} \cdots 
  \gamma_{\alpha_{15} \alpha_{16}}^{ a_{19} a_{20}}\,.
\end{align}
Clearly, all information about the structure of the effective action
terms is contained in these Levi-Civita gamma matrix
traces. The~$t_{16}$ tensor of course leads to amplitudes which can
equivalently be written using the well-known~$t_8 t_8$ product. Let us
therefore now turn to the computation of the other two traces.

\subsection{Evaluation of the Levi-Civita gamma traces}
\label{s:evaluation}

Having reduced the problem of computing vertex operator correlators to
the evaluation of Levi-Civita gamma matrix traces, one is looking for
an effective way to determine the tensor structure of the~$t_{18}$ and
$t_{20}$ tensors. We here follow a slight modification of the
procedure recently described in~\cite{Green:2005qr}.

\begin{figure}[th]
\begin{center}
\includegraphics[width=9cm]{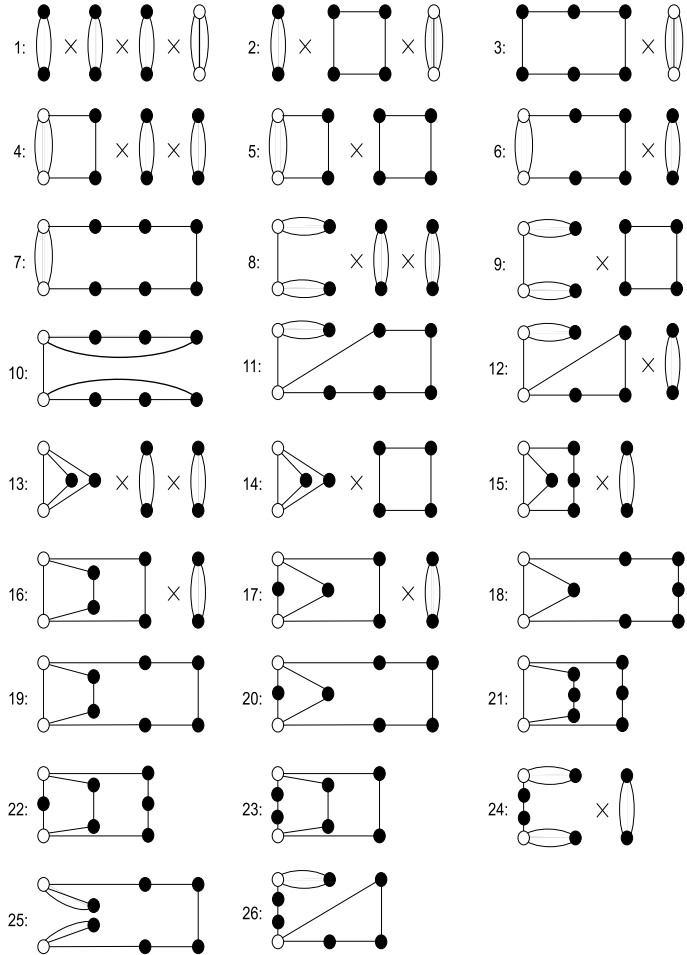}
\end{center}
\caption{The 26 graphs used to express the gamma trace~\eqref{e:t18},
  relevant for the $(DF)^2\,R^2$ terms in the amplitude. Black and white
  dots represent~$X^{ab}$ and~$Y^{abc}$ tensors
  respectively.\label{DF2R2Graphs} }
\end{figure}

The structure of the $t_{16}$, $t_{18}$ and $t_{20}$ tensors is most
conveniently written down in a form contracted with antisymmetric
dummy tensors~$Y^{abc}$ and~$X^{ab}$, such as to reveal the symmetry
of the $t$-tensors. The result has to be a linear combination of all
possible contractions of the $X$ and $Y$.\footnote{For completeness,
let us mention that in this notation the expression for the
contracted~$t_{16}$ tensor equals
\begin{multline}
t_{16}^{a_1 a_2\ldots a_{16}}\, X_{a_1a_2} \cdots X_{a_{17} a_{18}} = \\[1ex]
105\cdot 2^{19} \Big( -5\, \Tr(X^2)^4 +384\,\Tr(X^8) -256 \,\Tr(X^2) \Tr(X^6)
                                     +72\,\Tr(X^2)^2 \Tr(X^4) - 48\,\Tr(X^4)^2 \Big)\,.
\end{multline}}
For the~$t_{18}$ tensor the decomposition reads
\begin{equation}
\label{e:t18}
t_{18}^{a_1 a_2 \ldots a_{18}}Y_{a_1 a_2 a_3} Y_{a_4 a_5 a_6} X_{a_7 a_8}
\cdots X_{a_{17} a_{18}} = c_1\, Y^{abc}Y_{abc} (X^{de}X_{de})^3 + c_2\,
Y^{abc}Y_{abc} \Tr X^2\, \Tr X^4 + \ldots
\end{equation}
A simple group theory calculation shows that there are~26 possible
contractions for the~$t_{18}$ tensor: the tensor product
\begin{equation}
\left(\tableau{1 1 1}\right)^2_{\text{sym}} 
 \bigotimes \left(\tableau{1 1}\right)^6_{\text{sym}}
\end{equation}
contains 26 singlets in~SO(9). In figure~\ref{DF2R2Graphs} these
contractions have been visualised by representing~$X$ with a black dot
with 2 legs, $Y$~with a white dot with 3 legs and a contraction by
linking two legs. Repeatedly filling the components of~$X$ and $Y$
with random numbers and evaluating both the values of the~26 graphs
and the $t_{18}$ contraction (which can be done numerically in a
reasonable amount of time), one obtains an overdetermined linear
system of equations for the $c_1,\ldots,c_{26}$ and thus the tensor
structure of $t_{18}$, as shown in table~\ref{TensorStructureDF2R2} (an
alternative method to obtain the~$c_i$ coefficients, based on a
backtracking algorithm, was described in~\cite{Green:2005qr}).

\begin{table}[ht]
\center
\begin{tabular}{|c||c|c|}
\hline Scalar $i$ & Tensor structure & factor $c_{i}$ \\
\hline 1 & $\delta^{a_{1} a_{4}} \delta^{a_{2} a_{5}}
\delta^{a_{3} a_{6}} \delta^{a_{7} a_{9}} \delta^{a_{8} a_{10}}
\delta^{a_{11} a_{13}} \delta^{a_{12} a_{14}} \delta^{a_{15}
a_{17}} \delta^{a_{16} a_{18}}$ & 440 \\
2 & $ \delta^{a_{1} a_{4}} \delta^{a_{2} a_{5}} \delta^{a_{3}
a_{6}} \delta^{a_{7} a_{9}} \delta^{a_{8} a_{10}} \delta^{a_{11}
a_{13}} \delta^{a_{12} a_{15}} \delta^{a_{14}
a_{17}} \delta^{a_{16} a_{18}}$ & -2400 \\
3 & $ \delta^{a_{1} a_{4}} \delta^{a_{2} a_{5}} \delta^{a_{3}
a_{6}} \delta^{a_{7} a_{9}} \delta^{a_{8} a_{11}} \delta^{a_{10}
a_{13}} \delta^{a_{12} a_{15}} \delta^{a_{14}
a_{17}} \delta^{a_{16} a_{18}}$ & 2560\\
4 & $ \delta^{a_{1} a_{4}} \delta^{a_{2} a_{5}} \delta^{a_{3}
a_{7}} \delta^{a_{6} a_{9}} \delta^{a_{8} a_{10}} \delta^{a_{11}
a_{13}} \delta^{a_{12} a_{14}} \delta^{a_{15}
a_{17}} \delta^{a_{16} a_{18}}$ & -7200 \\
5 & $ \delta^{a_{1} a_{4}} \delta^{a_{2} a_{5}} \delta^{a_{3}
a_{7}} \delta^{a_{6} a_{9}} \delta^{a_{8} a_{10}} \delta^{a_{11}
a_{13}} \delta^{a_{12} a_{15}} \delta^{a_{14}
a_{17}} \delta^{a_{16} a_{18}}$ & 17280\\
6 & $ \delta^{a_{1} a_{4}} \delta^{a_{2} a_{5}} \delta^{a_{3}
a_{7}} \delta^{a_{6} a_{9}} \delta^{a_{8} a_{11}} \delta^{a_{10}
a_{13}} \delta^{a_{12} a_{14}} \delta^{a_{15}
a_{17}} \delta^{a_{16} a_{18}}$ & 23040\\
7 & $ \delta^{a_{1} a_{4}} \delta^{a_{2} a_{5}} \delta^{a_{3}
a_{7}} \delta^{a_{6} a_{9}} \delta^{a_{8} a_{11}} \delta^{a_{10}
a_{13}} \delta^{a_{12} a_{15}} \delta^{a_{14}
a_{17}} \delta^{a_{16} a_{18}}$ & -46080\\
8 & $ \delta^{a_{1} a_{4}} \delta^{a_{2} a_{7}} \delta^{a_{3}
a_{8}} \delta^{a_{5} a_{9}} \delta^{a_{6} a_{10}} \delta^{a_{11}
a_{13}} \delta^{a_{12} a_{14}} \delta^{a_{15}
a_{17}} \delta^{a_{16} a_{18}}$ & -720\\
9 & $ \delta^{a_{1} a_{4}} \delta^{a_{2} a_{7}} \delta^{a_{3}
a_{8}} \delta^{a_{5} a_{9}} \delta^{a_{6} a_{10}} \delta^{a_{11}
a_{13}} \delta^{a_{12} a_{15}} \delta^{a_{14}
a_{17}} \delta^{a_{16} a_{18}}$ & -2880\\
10 & $ \delta^{a_{1} a_{4}} \delta^{a_{2} a_{7}} \delta^{a_{3}
a_{9}} \delta^{a_{5} a_{11}} \delta^{a_{6} a_{13}} \delta^{a_{8}
a_{15}} \delta^{a_{10} a_{16}} \delta^{a_{12}
a_{17}} \delta^{a_{14} a_{18}}$ & -23040\\
11 & $ \delta^{a_{1} a_{4}} \delta^{a_{2} a_{7}} \delta^{a_{3}
a_{8}} \delta^{a_{5} a_{9}} \delta^{a_{6} a_{11}} \delta^{a_{10}
a_{13}} \delta^{a_{12} a_{15}} \delta^{a_{14}
a_{17}} \delta^{a_{16} a_{18}}$ & 0\\
12 & $ \delta^{a_{1} a_{4}} \delta^{a_{2} a_{7}} \delta^{a_{3}
a_{8}} \delta^{a_{5} a_{9}} \delta^{a_{6} a_{11}} \delta^{a_{10}
a_{13}} \delta^{a_{12} a_{14}} \delta^{a_{15}
a_{17}} \delta^{a_{16} a_{18}}$ & 11520\\
13 & $ \delta^{a_{1} a_{4}} \delta^{a_{2} a_{7}} \delta^{a_{3}
a_{9}} \delta^{a_{5} a_{8}} \delta^{a_{6} a_{10}} \delta^{a_{11}
a_{13}} \delta^{a_{12} a_{14}} \delta^{a_{15}
a_{17}} \delta^{a_{16} a_{18}}$ & 4320\\
14 & $ \delta^{a_{1} a_{4}} \delta^{a_{2} a_{7}} \delta^{a_{3}
a_{9}} \delta^{a_{5} a_{8}} \delta^{a_{6} a_{10}} \delta^{a_{11}
a_{13}} \delta^{a_{12} a_{15}} \delta^{a_{14}
a_{17}} \delta^{a_{16} a_{18}}$ & -5760\\
15 & $ \delta^{a_{1} a_{4}} \delta^{a_{2} a_{7}} \delta^{a_{3}
a_{9}} \delta^{a_{5} a_{8}} \delta^{a_{6} a_{11}} \delta^{a_{10}
a_{13}} \delta^{a_{12} a_{14}} \delta^{a_{15}
a_{17}} \delta^{a_{16} a_{18}}$ & -46080\\
16 & $ \delta^{a_{1} a_{4}} \delta^{a_{2} a_{7}} \delta^{a_{3}
a_{9}} \delta^{a_{5} a_{11}} \delta^{a_{6} a_{13}} \delta^{a_{8}
a_{12}} \delta^{a_{10} a_{14}} \delta^{a_{15}
a_{17}} \delta^{a_{16} a_{18}}$ & 23040\\
17 & $ \delta^{a_{1} a_{7}} \delta^{a_{2} a_{9}} \delta^{a_{3}
a_{11}} \delta^{a_{4} a_{8}} \delta^{a_{5} a_{10}} \delta^{a_{6}
a_{13}} \delta^{a_{12} a_{14}} \delta^{a_{15}
a_{17}} \delta^{a_{16} a_{18}}$ & -11520\\
18 & $ \delta^{a_{1} a_{4}} \delta^{a_{2} a_{7}} \delta^{a_{3}
a_{9}} \delta^{a_{5} a_{8}} \delta^{a_{6} a_{11}} \delta^{a_{10}
a_{13}} \delta^{a_{12} a_{15}} \delta^{a_{14}
a_{17}} \delta^{a_{16} a_{18}}$ & 92160\\
19 & $ \delta^{a_{1} a_{4}} \delta^{a_{2} a_{7}} \delta^{a_{3}
a_{9}} \delta^{a_{5} a_{11}} \delta^{a_{6} a_{13}} \delta^{a_{8}
a_{12}} \delta^{a_{10} a_{15}} \delta^{a_{14}
a_{17}} \delta^{a_{16} a_{18}}$ & -92160\\
20 & $ \delta^{a_{1} a_{7}} \delta^{a_{2} a_{9}} \delta^{a_{3}
a_{11}} \delta^{a_{4} a_{8}} \delta^{a_{5} a_{10}} \delta^{a_{6}
a_{13}} \delta^{a_{12} a_{15}} \delta^{a_{14}
a_{17}} \delta^{a_{16} a_{18}}$ & 0\\
21 & $ \delta^{a_{1} a_{4}} \delta^{a_{2} a_{7}} \delta^{a_{3}
a_{9}} \delta^{a_{5} a_{11}} \delta^{a_{6} a_{13}} \delta^{a_{8}
a_{15}} \delta^{a_{10} a_{17}} \delta^{a_{12}
a_{16}} \delta^{a_{14} a_{18}}$ & 46080\\
22 & $ \delta^{a_{1} a_{7}} \delta^{a_{2} a_{9}} \delta^{a_{3}
a_{11}} \delta^{a_{4} a_{8}} \delta^{a_{5} a_{13}}
\delta^{a_{6} a_{15}} \delta^{a_{10} a_{14}} \delta^{a_{12}
a_{17}} \delta^{a_{16} a_{18}}$ & 92160\\
23 & $ \delta^{a_{1} a_{7}} \delta^{a_{2} a_{9}} \delta^{a_{3}
a_{11}} \delta^{a_{4} a_{13}} \delta^{a_{5} a_{15}} \delta^{a_{6}
a_{17}} \delta^{a_{8} a_{14}} \delta^{a_{10}
a_{16}} \delta^{a_{12} a_{18}}$ & 0\\
24 & $ \delta^{a_{1} a_{7}} \delta^{a_{2} a_{8}} \delta^{a_{3}
a_{9}} \delta^{a_{4} a_{11}} \delta^{a_{5} a_{12}} \delta^{a_{6}
a_{13}} \delta^{a_{10} a_{14}} \delta^{a_{15}
a_{17}} \delta^{a_{16} a_{18}}$ & 5760\\
25 & $ \delta^{a_{1} a_{7}} \delta^{a_{2} a_{8}} \delta^{a_{3}
a_{9}} \delta^{a_{4} a_{11}} \delta^{a_{5} a_{12}} \delta^{a_{6}
a_{13}} \delta^{a_{10} a_{15}} \delta^{a_{14}
a_{17}} \delta^{a_{16} a_{18}}$ & 0\\
26 & $ \delta^{a_{1} a_{7}} \delta^{a_{2} a_{8}} \delta^{a_{3}
a_{9}} \delta^{a_{4} a_{11}} \delta^{a_{5} a_{13}} \delta^{a_{6}
a_{15}} \delta^{a_{10} a_{12}} \delta^{a_{14}
a_{17}} \delta^{a_{16} a_{18}}$ & -46080\\
\hline
\end{tabular}
\caption{Decomposition of the $t_{18}$ tensor, defined
in~\eqref{e:t18def}, in terms of Lorentz
singlets.\label{TensorStructureDF2R2}}
\end{table}

For the $t_{20}$ one proceeds in a similar way. After contraction with
the~$X$ and~$Y$ tensors we end up with
\begin{equation}
t_{20}^{a_1 a_2 \ldots a_{20}}Y_{a_1 a_2 a_3} \cdots Y_{a_{10} a_{11}
  a_{12}} X_{a_{13} a_{14}} \cdots X_{a_{19} a_{20}} = c_1\, Y^{abc}Y_{abc}
Y^{def}Y_{def} (X^{gh}X_{gh})^2 + \ldots\,.
\end{equation}
In this case there are 83 possible contractions of the~$X$ and~$Y$
tensors, which follows from the fact that the tensor product
\begin{equation}
\label{e:t20tensorprod}
\left(\tableau{1 1 1}\right)^4_{\text{sym}} 
  \bigotimes \left(\tableau{1 1}\right)^4_{\text{sym}}
\end{equation}
contains 83 singlets in~SO(9). Because the procedure that leads to
the determination of the~83 Lorentz singlets and their coefficients is
entirely the same as for the~$t_{18}$ tensor, we will refrain from
spelling out these results. 

There is, however, one subtle point which is worth mentioning. We are
computing the membrane vertex operator correlators in the light-cone
gauge, i.e.~the indices take values in~SO(9). However, when the tensor
product~\eqref{e:t20tensorprod} is computed in~SO(11), one finds one
additional singlet. Therefore, when the~84 basis elements in SO(11)
are reduced to SO(9), one has to find one non-trivial identity. When,
subsequently, the tensors $X$ and $Y$ are replaced with~$DF$ tensors
(see the next subsection), this identity leads to one non-trivial
identity between the various elements in the basis of~$(DF)^4$
invariants.  This identity is spelled out in~\eqref{zeroIdentity}.  It
implies that the light-cone computation will always leave one linear
combination of terms in the covariant action undetermined.  However,
as we will see below, a covariant computation of the four-point
amplitude resulting from this linear combination yields a vanishing
result. While the ambiguity thus restricts our ability to determine
the full effective action, it is irrelevant for the covariantisation
of our light-cone amplitudes.

\subsection{Obtaining the amplitude}

The final step towards the amplitude is replacing $XX \rightarrow R$,
$XY \rightarrow DF$ and adequately symmetrising, e.g.~for the
$(DF)^{2}\,R^2$ amplitude
\begin{multline}
Y_{a_1 a_2 a_3} Y_{a_4 a_5 a_6} X_{a_7 a_8} X_{a_9 a_{10}} X_{a_{11}
  a_{12}} X_{a_{13} a_{14}} X_{a_{15} a_{16}} X_{a_{17} a_{18}}\\[1ex] 
\rightarrow \left(D_{[a_7}F_{a_{8}] a_1 a_2 a_3} D_{[a_{9}} F_{a_{10}] a_4 a_5 a_6} R_{a_{11} a_{12} a_{13} a_{14}} R_{a_{15} a_{16} a_{17} a_{18}}\right)_{\text{sym}}\,.
\end{multline}
Here the suffix ``sym'' denotes symmetrisation in the 6 index pairs
$[a_7 a_8]$,\ldots,$[a_{17} a_{18}]$ and the two index triplets $[a_1 a_2
a_3]$, $[a_4 a_5 a_6]$. The amplitude is then simply
\begin{multline} 
{\cal A}_{(DF)^2 R^2} = 
 \big(c_1\, \delta^{a_1 a_4} \delta^{a_2 a_5} \delta^{a_3 a_6} 
            \delta^{a_7 a_9}\delta^{a_8 a_{10}} \delta^{a_{11} a_{13}}
            \delta^{a_{12} a_{14}} \delta^{a_{15} a_{17}}\delta^{a_{16} a_{18}} 
   + \ldots\big) \\[1ex]
\times \big( D_{[a_7}F_{a_{8}] a_1 a_2 a_3} D_{[a_{9}} F_{a_{10}] a_4 a_5 a_6} 
         R_{a_{11} a_{12} a_{13} a_{14}} R_{a_{15} a_{16} a_{17} a_{18}}\big)_{\text{sym}}. 
\label{eqn_originalAmplitude}
\end{multline}
Eventually, one wants to bring the amplitude in an appealing form,
writing it in terms of a (suitably chosen) basis. To do so, one has to
use, besides the simple monoterm symmetries, multiterm symmetries,
which can be tackled with the method of Young
projectors~\cite{Green:2005qr}. A nice consequence of our method of
calculation is that it leads straightaway to a covariant form of the
action, by virtue of the fact that the vertex operators are written in
terms of linearised Riemann tensors and linearised derivatives of
field strengths.

\section{Results}
\subsection{The amplitudes and the effective action}

Using the method outlined in the previous section, the
two-graviton/two-threeform and four-threeform amplitudes can now be
computed. Summarising, the four-particle amplitudes are then given by
the following expressions (in terms of the Fulling
basis~\cite{Fulling:1992vm,Peeters:2000qj} and the tensor monomial
basis given in the appendix):
\begin{multline}
\label{e:R4}
{\cal A}_{R^4} = 2^{20} \big( -192\,A_4 + 768\,A_7\big)\,,\hfill
\end{multline}
\begin{multline} 
\label{e:AR2DF2}
{\cal A}_{(DF)^2\, R^2} = 2^{21} \cdot 3^{3} 
\big( -24\bA_{5} -48\bA_{8} -24\bA_{10} -6\bA_{12} -12\bA_{13} +12\bA_{14}\\[1ex]
 +8\bA_{16} -4 \bA_{20} + \bA_{22} + 4\bA_{23} + \bA_{24} \big) \,,
\end{multline}
\begin{multline} 
\label{e:ADF4}
{\cal A}_{(DF)^4} = 2^{19} \cdot 3^{7} 
\big( 3 \bB_5 + \bB_6 - 9\bB_8 + \bB_9 - 72\bB_{12} + 9\bB_{14} + 18 \bB_{17} - 9
\bB_{18} - 72\bB_{19} - \bB_{22}\big)\,.
\end{multline}
We should emphasise that these expressions are amplitudes, even though
we have expressed them in terms of the effective action contact terms
which can produce them. As explained in the appendix, there are 6
linear combinations of $(DF)^2\,R^2$ terms and 9 linear combinations of
$(DF)^4$ terms in the effective action which lead to a vanishing
amplitude. Their coefficients are therefore not determined by the
expressions above. This is similar to the well-known fact that the term
\begin{equation}
\int\!{\rm d}^{10}x\, \epsilon_{mn r_1\ldots r_8} \epsilon_{mn
  s_1\ldots s_8} R_{r_1 r_2 s_1 s_2}\cdots R_{r_7 r_8 s_7 s_8}
\end{equation}
in the string effective action leads to a vanishing four-graviton
amplitude. In order to determine its coefficient, a computation of the
five-graviton amplitude is required. 

The covariant effective action for the~$(DF)^2\, R^2$ and the~$(DF)^4$
terms is thus given by~\eqref{e:AR2DF2} and~\eqref{e:ADF4} (with
the~$\bA_i$ and $\bB_i$ now interpreted as covariant terms in the
action), plus an undetermined linear sum of the~$Z_i$
and~$\tilde{Z}_i$ combinations given in~\eqref{e:Zs}
and~\eqref{e:Ztildes}\footnote{The~$\tilde{Z}_i$ terms also contain
the SO(9) identity discussed at the end of section~\ref{s:evaluation},
so that no further ambiguity is introduced by this
identity.}. Unfortunately the number of undetermined coefficients in
the action is thus rather large (6 and~9 coefficients respectively)
and determining these requires information from higher-point
amplitudes. We will not attempt that here (note, however, that a
similar ambiguity is also present in the superstring effective action,
where e.g.~the~$(DH)^2 R^2$ terms are determined by a four-point
calculation only up to a four-parameter family of
terms~\cite{Peeters:2003pv,Peeters:2001ub}).

Finally, we have also verified that when our one-loop amplitudes are
divided by the product of the Mandelstam variables~$stu$, they become
identical to the tree-level exchange amplitudes with the same external
particles. This lends support to the method followed by Deser \&
Seminara for the construction of higher-derivative
counterterms~\cite{Deser:1998jz,Deser:2000xz,Deser:2005kb}. While we
agree on the form of the $(DF)^2\, R^2$ effective action, the
local~$(DF)^4$ action obtained in~\cite{Deser:2000xz} yields an
amplitude which differs from the one obtained from our
expression~\eqref{e:ADF4}. We do, however, agree
with~\cite{Deser:2000xz} on the original, nonlocal amplitude.

\subsection{Cross check via compactification}

Upon compactification of our eleven-dimensional results to ten
dimensions, it should be possible to match our amplitudes with known
expressions for the two-graviton/two-twoform and four-twoform
amplitudes in type-IIA string theory. These were computed a long time
ago~\cite{Gross:1987mw}. Writing the eleven-dimensional indices as
$A=(a,11)$, the compactification rule is
\begin{equation}
F_{11\, abc} = H_{abc}\,.
\end{equation}
The string computations of~\cite{Gross:1987mw}, which were done in the
Ramond-Neveu-Schwarz formalism, show that the~$(DH)^2\, R^2$ and $(DH)^4$
terms in the effective action are obtained from the~$R^4$ action by
shifting the spin connection with the curvature of the two-form gauge
field (the torsion),
\begin{equation}
\label{e:shifted}
\omega \rightarrow \omega + H_{(3)}\qquad\rightarrow\qquad
R_{abcd} \rightarrow R_{abcd} + D_{[a} H_{b]cd}\,.
\end{equation}
The amplitudes are thus simply obtained by replacing an appropriate
number of Riemann tensors with derivatives of the gauge field
curvature in the expression
\begin{equation} 
\label{e:t8t8R4}
{\cal A}_{R^4} = t_8^{a_1 \cdots a_8} t_8^{b_1 \cdots b_8}\, R_{a_1 a_2 b_1 b_2}\,
  R_{a_3 a_4 b_3 b_4}\, R_{a_5 a_6 b_5 b_6} \, R_{a_7 a_8 b_7 b_8}\,.
\end{equation}
Note that due to the fact that $D_{[a} H_{b]cd}$ has a different
symmetry structure compared to $R_{abcd}$,
\begin{equation}
R_{abcd} = R_{cdab} \quad \mbox{vs.} \quad D_{[a} H_{b]cd} = -D_{[c} H_{d]dab}
\end{equation}
one must not first perform the contractions in~\eqref{e:t8t8R4} (see
e.g.~\cite{Peeters:2000qj} for the result) and then perform the
substitution~\eqref{e:shifted}. Instead, the substitution must be
made at the level of~\eqref{e:t8t8R4} directly.

As it should be, we now find that when we compactify our $(DF)^2\,R^2$
and $(DF)^4$ amplitudes, they precisely match the corresponding
$(DH)^2\,R^2$ resp.~$(DH)^4$ amplitudes of~\cite{Gross:1987mw}. This
match of the compactified action provides a strong consistency check
on our computations.

\vfill\eject
\section{Discussion}

We have computed the~$(DF)^2 R^2$ and $(DF)^4$ terms in the M-theory
effective action using a light-cone gauge superparticle or
supermembrane calculation. Let us conclude by discussing some open
issues and possible applications of our results.

In order to compute terms in the effective action which do not contain
derivatives on the gauge fields, like~$R^3 F^2$, it is necessary to
analyse higher-point amplitudes. In this case, it becomes possible to
saturate the fermionic zero modes without restricting to the
superparticle terms~\eqref{e:Vsuperparticle} in the supermembrane
vertex operators. At first sight, this seems to imply that one needs
full control over the bosonic correlators, which at least for the time
being is not sufficiently understood for the supermembrane. However,
it may be that certain simplifications occur because of the fact that
e.g.~the $(DF)^2\,R^2$ and $F^2\,R^3$ terms in the effective action
are related to each other by nonlinear supersymmetry. A similar
supersymmetry relation is, presumably, responsible for the fact that
the insertion point integrals of string five-point amplitudes reduce
to extremely simple expressions~\cite{Peeters:2003pv}.

One of our motivations for the computation presented here is given by
recent results of Damour and Nicolai~\cite{Damour:2005zb}. Their work
is concerned with the study of dynamics near space-like singularities
in eleven dimensions, following the seminal work of Belinsky,
Khalatnikov and Lifshitz~\cite{Belinsky:1970ew,Belinsky:1982pk}. In
the late-time limit, it has been known for some
time~\cite{Damour:2002et} that supergravity reduces to a point
particle sigma model on the infinite dimensional coset
space~$E_{10}/K(E_{10})$. It was shown in~\cite{Damour:2005zb} that
the higher-derivative~$R^4$ terms in the M-theory effective action are
also encoded, in an intriguing way, in the structure of the~$E_{10}$
root space. It would be very interesting to analyse whether the~$(DF)^2
R^2$ and~$(DF)^4$ terms agree in a similar way with the structure
expected from~$E_{10}$. Such a match would provide strong support
for an entirely new perspective on the construction of the M-theory
effective action.

\section*{Acknowledgements}

We thank Stanley Deser for comments on a draft of this paper and for
correspondence concerning the results of~\cite{Deser:2000xz}.

\vfill\eject
\appendix
\section{Appendix}
\subsection{The $(DF)^2\,R^2$ basis}

If we impose the linearised lowest-order equations of
motion\footnote{This we can do, as field redefinitions of $g_{ab}$ and
$C_{abc}$ allow us to always remove terms proportional to the equations
of motion at the first level of higher derivative corrections.},
there are 24 possible~$(DF)^2\,R^2$ terms in the action, which follows
from the fact that the tensor product
\begin{equation}
\left(\tableau{2 2}\right)^2_{\text{sym}} \bigotimes \left(\tableau{2 1 1
  1}\right)^2_{\text{sym}}
\end{equation}
contains 24 singlets in SO(11). We choose a basis given by
\begin{equation}
\begin{aligned}
\bA_{1}  &=  R_{abcd} R_{efgh}             D^{e}F^{agh}{}_{i}    D^{c}F^{bdfi} \hspace{2cm} &
\bA_{13} &= R_{abcd} R_{e}{}^{a}{}_{f}{}^{c} D_{i}F^{bf}{}_{gh} D^{i}F^{degh}\\
\bA_{2}  &=  R_{abcd} R_{efgh}             D^{e}F^{acg}{}_{i}    D^{h}F^{bdfi}\hspace{2cm} &
\bA_{14} &= R_{abcd} R_{e}{}^{a}{}_{f}{}^{c} D_{i}F^{bd}{}_{gh} D^{i}F^{efgh}\\
\bA_{3}  &=  R_{abcd} R_{efgh}             D^{e}F^{acg}{}_{i}    D^{f}F^{bdhi}\hspace{2cm} &
\bA_{15} &= R_{abcd} R_{e}{}^{a}{}_{f}{}^{c} D^{b}F^{f}{}_{ghi} D^{e}F^{dghi}\\
\bA_{4}  &=  R_{abcd} R_{efgh}             D_{i}F^{cdgh}         D^{f}F^{iabe}\hspace{2cm} &
\bA_{16} &= R_{abcd} R_{e}{}^{a}{}_{f}{}^{c} D^{b}F^{d}{}_{ghi} D^{e}F^{fghi}\\
\bA_{5}  &=  R_{abcd} R_{efg}{}^{d}        D^{a}F^{bc}{}_{hi}    D^{e}F^{fghi}\hspace{2cm} &
\bA_{17} &= R_{abcd} R_{e}{}^{a}{}_{f}{}^{c} D^{b}F^{e}{}_{ghi} D^{d}F^{fghi}\\
\bA_{6}  &=  R_{abcd} R_{efg}{}^{d}        D^{a}F^{be}{}_{hi}    D^{c}F^{fghi}\hspace{2cm} &
\bA_{18} &= R_{abcd} R_{e}{}^{a}{}_{f}{}^{c} D_{i}F^{ef}{}_{gh} D^{d}F^{bghi}\\
\bA_{7}  &=  R_{abcd} R_{efg}{}^{d}        D^{a}F^{be}{}_{hi}    D^{g}F^{cfhi}\hspace{2cm} &
\bA_{19} &= R_{abcd} R_{ef}{}^{cd}        D_{i}F^{ae}{}_{gh}    D^{i}F^{bfgh}\\
\bA_{8}  &=  R_{abcd} R_{efg}{}^{d}        D^{a}F^{ce}{}_{hi}    D^{b}F^{fghi}\hspace{2cm} &
\bA_{20} &= R_{abcd} R_{ef}{}^{cd}        D^{a}F^{e}{}_{ghi}    D^{b}F^{fghi}\\
\bA_{9}  &=  R_{abcd} R_{efg}{}^{d}        D^{a}F^{ce}{}_{hi}    D^{f}F^{bghi}\hspace{2cm} &
\bA_{21} &= R_{abcd} R_{ef}{}^{cd}        D^{a}F^{e}{}_{ghi}    D^{f}F^{bghi}\\
\bA_{10} &= R_{abcd} R_{efg}{}^{d}        D_{i}F^{cegh}         D^{i}F^{abfh}\hspace{2cm} &
\bA_{22} &= R_{abcd} R_{e}{}^{acd}        D^{b}F_{fghi}         D^{e}F^{fghi}\\
\bA_{11} &= R_{abcd} R_{efg}{}^{d}        D_{h}F^{abf}{}_{i}    D^{i}F^{cegh}\hspace{2cm} &
\bA_{23} &= R_{abcd} R_{e}{}^{acd}        D_{i}F^{b}{}_{fgh}    D^{i}F^{efgh}\\
\bA_{12} &= R_{abcd} R_{efg}{}^{d}        D^{c}F^{ef}{}_{hi}    D^{g}F^{bahi}\hspace{2cm} &
\bA_{24} &= R_{abcd} R^{abcd}             D_{e}F_{fghi}         D^{f}F^{eghi}
\end{aligned}
\end{equation}
The four-point amplitudes resulting from these contact terms are,
however, not all independent. We find that the following~6 linear
combinations lead to a vanishing four-point amplitude:
\begin{equation}
\label{e:Zs}
\begin{aligned}
Z_1 &= 48\bA_1 + 48\bA_2 - 48\bA_3 + 36\bA_4 + 96\bA_6 + 48\bA_7 - 48\bA_8 +
96\bA_{10} \\[1ex]
&\quad\quad + 12\bA_{12} +24\bA_{13} -12\bA_{14} + 8\bA_{15} + 8\bA_{16} - 16\bA_{17} + 6\bA_{19} + 2\bA_{22} + \bA_{24}\,,\\[1ex]
Z_2 &=-48\bA_1 -48\bA_2 -24\bA_4 -24\bA_5 +48\bA_6 -48\bA_8 -24\bA_9
-72\bA_{10}\\[1ex]
&\quad\quad -24\bA_{13} +24\bA_{14} -\bA_{22} +4\bA_{23}\,,\\[1ex]
Z_3 &= 12\bA_1 + 12\bA_2 - 24\bA_3 + 9\bA_4 +48\bA_6 + 24\bA_7 - 24\bA_8 + 24\bA_{10}
+ 6\bA_{12} + 6\bA_{13}\\[1ex]
&\quad\quad + 4\bA_{15} - 4\bA_{17} + 3\bA_{19} + 2\bA_{21}\,,\\[1ex]
Z_4 &= 12\bA_1 + 12\bA_2 - 12\bA_3 + 9\bA_4 +24\bA_6 + 12\bA_7 - 12\bA_8 + 24\bA_{10}
+ 3\bA_{12} \\[1ex]
&\quad\quad + 6\bA_{13} + 4\bA_{15} - 4\bA_{17} + 2\bA_{20}\,,\\[1ex]
Z_5 &= 4\bA_{3} -8\bA_{6} -4\bA_7 + 4\bA_8 -\bA_{12} -2\bA_{14} + 4\bA_{18}\,,\\[1ex]
Z_6 &= \bA_4 + 2\bA_{11} \,.
\end{aligned}
\end{equation}
In order to determine the coefficients of these linear combinations of
terms in the effective action it is necessary to consider
amplitudes with more external legs (which we will not attempt in the
present paper).

\subsection{The $(DF)^4$ basis}

Just like the basis of~$(DF)^2\,R^2$ terms, the basis for the~$(DF)^4$
terms consists, at least at linearised on-shell level, of 24 elements:
the tensor product
\begin{equation}
\label{e:DF4tens}
\left(\tableau{2 1 1 1}\right)^4_{\text{sym}}
\end{equation}
contains 24 singlets in SO(11). We choose these to be
\begin{equation}
\begin{aligned}
\bB_{1}  &=  D_{a}F_{bcde} D^{a}F^{bcde}      D_{f}F_{ghij}      D^{f}F^{ghij}\hspace{1.5cm} &
\bB_{13} &= D_{a}F_{bcde} D^{a}F^{b}{}_{fgh} D^{c}F^{fg}{}_{ij} D^{d}F^{ehij}\\
\bB_{2}  &=  D_{a}F_{bcde} D^{a}F^{bcd}{}_{f} D^{e}F_{ghij}      D^{f}F^{ghij}\hspace{1.5cm} &
\bB_{14} &= D_{a}F_{bcde} D^{a}F^{b}{}_{fgh} D_{i}F^{cdf}{}_{j} D^{i}F^{eghj}\\
\bB_{3}  &=  D_{a}F_{bcde} D^{a}F^{bcd}{}_{f} D_{g}F^{e}{}_{hij} D^{g}F^{fhij}\hspace{1.5cm} &
\bB_{15} &= D_{a}F_{bcde} D^{a}F_{fghi}      D^{b}F^{cdf}{}_{j} D^{g}F^{ehij}\\
\bB_{4}  &=  D_{a}F_{bcde} D^{a}F^{bc}{}_{fg} D^{d}F^{e}{}_{hij} D^{f}F^{ghij}\hspace{1.5cm} &
\bB_{16} &= D_{a}F_{bcde} D^{a}F_{fghi}      D^{b}F^{fgh}{}_{j} D^{i}F^{cdej}\\
\bB_{5}  &=  D_{a}F_{bcde} D^{a}F^{b}{}_{fgh} D_{i}F^{cde}{}_{j} D^{f}F^{ghij}\hspace{1.5cm} &
\bB_{17} &= D_{a}F_{bcde} D^{b}F^{ac}{}_{fg} D_{h}F^{df}{}_{ij} D^{i}F^{eghj}\\
\bB_{6}  &=  D_{a}F_{bcde} D^{a}F^{b}{}_{fgh} D_{i}F^{cde}{}_{j} D^{j}F^{fghi}\hspace{1.5cm} &
\bB_{18} &= D_{a}F_{bcde} D^{b}F^{a}{}_{fgh} D_{i}F^{cdf}{}_{j} D^{j}F^{eghi}\\
\bB_{7}  &=  D_{a}F_{bcde} D^{b}F^{ac}{}_{fg} D^{d}F^{e}{}_{hij} D^{h}F^{fgij}\hspace{1.5cm} &
\bB_{19} &= D_{a}F_{bcde} D^{b}F^{c}{}_{fgh} D^{f}F^{dg}{}_{ij} D^{i}F^{aehj}\\
\bB_{8}  &=  D_{a}F_{bcde} D^{b}F^{ac}{}_{fg} D_{h}F^{de}{}_{ij} D^{i}F^{fghj}\hspace{1.5cm} &
\bB_{20} &= D_{a}F_{bcde} D^{b}F^{c}{}_{fgh} D^{f}F^{ad}{}_{ij} D^{i}F^{eghj}\\
\bB_{9}  &=  D_{a}F_{bcde} D^{b}F_{fghi}      D^{f}F^{ghi}{}_{j} D^{j}F^{acde}\hspace{1.5cm} &
\bB_{21} &= D_{a}F_{bcde} D^{b}F^{c}{}_{fgh} D^{d}F^{fg}{}_{ij} D^{h}F^{aeij}\\
\bB_{10} &= D_{a}F_{bcde} D^{a}F^{bc}{}_{fg} D_{h}F^{df}{}_{ij} D^{i}F^{eghj}\hspace{1.5cm} &
\bB_{22} &= D_{a}F_{bcde} D^{b}F_{fghi}      D^{f}F^{cde}{}_{j} D^{j}F^{aghi}\\
\bB_{11} &= D_{a}F_{bcde} D^{a}F^{bc}{}_{fg} D^{d}F^{f}{}_{hij} D^{g}F^{ehij}\hspace{1.5cm} &
\bB_{23} &= D_{a}F_{bcde} D^{b}F_{fghi}      D^{f}F^{cdg}{}_{j} D^{j}F^{aehi}\\
\bB_{12} &= D_{a}F_{bcde} D^{a}F^{b}{}_{fgh} D^{c}F^{df}{}_{ij} D^{i}F^{eghj}\hspace{1.5cm} &
\bB_{24} &= D_{a}F_{bcde} D^{b}F_{fghi}      D^{f}F^{acd}{}_{j} D^{g}F^{ehij}.
\end{aligned}
\end{equation}
However, when the indices are restricted to the transversal SO(9)
sector, the tensor product~\eqref{e:DF4tens} only contains 23
singlets. The SO(9) identity which relates the 24th basis element to
the others can be found by using the fact that~$\epsilon_{10}$ is
always zero in SO(9). This implies that
\begin{multline} 
0 = \epsilon^{a_1\cdots a_{10}} \epsilon^{b_1\cdots b_{10}} D_{a_1}F_{b_1 b_2 b_3 b_4}  D_{a_2}F_{a_3 b_5 b_6 b_7} D_{a_4} F_{a_5 a_6 b_8 b_9} D_{a_7}F_{a_8 a_9 a_{10} b_{10}} \\[1ex]
= \text{const} \cdot (-\bB_{1} + 8 \bB_{2} +16\bB_3 -96\bB_4 -32\bB_6 +144\bB_8 -16\bB_9 +96\bB_{11} + 1728 \bB_{12} -288 \bB_{13}\\[1ex]
 -144 \bB_{14} -32 \bB_{16} -576 \bB_{17} +288 \bB_{18} +1728 \bB_{19} -144 \bB_{21} +144 \bB_{23} ) + \mbox{off shell terms}\,. \label{zeroIdentity}
\end{multline}
The four-point amplitudes which are generated by the~$\bB_i$ terms are
again not all independent. It turns out that 9~linear combinations of
basis elements lead to a vanishing amplitude. These combinations are
given by
\begin{equation}
\label{e:Ztildes}
\begin{aligned}
\tilde{Z}_1 &= -\bB_3 + 12\bB_4 - 6\bB_5 + 72\bB_7 - 9\bB_8 -\bB_{9} + 54\bB_{10}
- 6\bB_{11} - 144\bB_{12} + 18\bB_{14} -27\bB_{18} +
18\bB_{21}\,,\\
\tilde{Z}_2 &= \bB_3 - 6\bB_5 -18\bB_7 +9\bB_8 +\bB_9 +6\bB_{11} + 9\bB_{18} + 18\bB_{23}\,,\\
\tilde{Z}_3 &= \bB_1 + 96\bB_{4} - 96\bB_5 + 32\bB_6 + 288\bB_7 + 64\bB_9 + 32\bB_{22}\,,\\
\tilde{Z}_4 &= -\bB_{10} + 2\bB_{12} + 2\bB_{20}\,,\\
\tilde{Z}_5 &= \bB_7 + \bB_{10} + 4\bB_{19}\,,\\
\tilde{Z}_6 &= -\bB_7 -\bB_{10} + 2\bB_{17}\,,\\
\tilde{Z}_7 &= \bB_1 - 8\bB_2 + 32\bB_6 + 32\bB_9 + 32\bB_{16}\,,\\
\tilde{Z}_8 &= -\bB_2 -12\bB_4 + 12\bB_5 -4\bB_9 - 12\bB_{11} + 36\bB_{15}\,,\\
\tilde{Z}_9 &= \bB_{10} - 2\bB_{12} + \bB_{13}\,.
\end{aligned}
\end{equation}
Note that the SO(9) identity~\eqref{zeroIdentity} is automatically
included in these vanishing relations.

\setlength{\bibsep}{5pt}

\begingroup\raggedright\endgroup

\end{document}